\documentclass[conference]{IEEEtran}
\IEEEoverridecommandlockouts
\usepackage{cite}
\usepackage{amsmath,amssymb,amsfonts}
\usepackage{algorithmic}
\usepackage{graphicx}
\usepackage{textcomp}
\usepackage{xcolor}
\usepackage{fancyhdr}
\usepackage{multirow}

\def\BibTeX{{\rm B\kern-.05em{\sc i\kern-.025em b}\kern-.08em
    T\kern-.1667em\lower.7ex\hbox{E}\kern-.125emX}}
\begin{document}

\title{Automated False Positive Filtering for esNetwork Alerts*\\
{\footnotesize \textsuperscript{*}Project report for internship at eSentire Inc.}
}

\author{\IEEEauthorblockN{Guangyi Zhu}
\IEEEauthorblockA{\textit{School of Computer Science} \\
\textit{University of Guelph}\\
Guelph, ON, Canada \\
guangyi@uoguelph.ca}
}

\maketitle

\begin{abstract}
An Intrusion Detection System (IDS) is one of the security tools that can automatically analyze network traffic and detect suspicious activities. They are widely implemented as security guarantee tools in various business networks. However, the high rate of false-positive alerts creates an overwhelming number of unnecessary alerts for security analysts to sift through. The esNetwork is an IDS product by eSentire Inc. This project focuses on reducing the false-positive alerts generated by esNetwork with the help of a Random Forest (RF) classifier. The RF model was built to classify the alerts as high and low and only pass high likelihood alerts to the analysts. As a result of evaluation experiments, this model can achieve an accuracy of 97\% for training validation, 88\% for testing with the recent data, and 58\% with Security Operation Centre (SOC) reviewed events. The evaluation result of the proposed model is intermediate because of the deficiency of clearly labeled data for training as well as the SOC-reviewed events for evaluation. The model still needs time to be fine-tuned to meet the industry deployment requirement.
\end{abstract}


\section{Introduction}
\subsection{Intrusion Detection System (IDS)}
IDSs are known for a strong method of defense against suspicious software and malicious network activities \cite{a1}. They were designed to provide alerts of high quality. There are several open-source IDS engines available, such as Suricata and Snort. However, unfortunately, they have also brought a problem of producing a high rate of false-positive alerts. The huge number of false-positive alerts can become a potential workload and can easily overwhelm Security Operation Centre (SOC) analysts. Pietraszek and Tanner\cite{1} have discussed the reasons. They summarized that the reasons include: 
\begin{enumerate}
    \item \textbf{Runtime Limitations}: the IDS/IPS cannot analyze all contents due to the harsh real-time requirement.
    \item \textbf{Specificity of detection signatures}: it is difficult to determine whether a signature is normal or not.
    \item \textbf{Dependency on the environment}: Some benign signatures may be malicious under specific circumstances.
    \item \textbf{Base-rate fallacy}: false-positive alerts are unavoidable.
\end{enumerate}

 With the increasing need for IDS, deducing the occurrence of false alerts has remained a common topic in the field for several years.
 
 \subsection{The esNetwork}
 
 The esNetwork by eSentire Inc. is a zero-latency IDS/IPS as a core component of Managed Detection and Response™ (MDR) service\footnote{$https://esentire-dot-com-assets.s3.ca-central-1.amazonaws.com/assets/resourcefiles/esNETWORK\_2019.pdf$}. The esNetwork sensors are deployed to capture runtime network traffic packets. By detecting and identifying the potential threats with threat intelligence analysis modules, the esNetwork can interrupt connections and escalate alerts to the SOC analysts for further investigations if necessary.
 
 \subsection{Project Objectives}
 
 The ATA Team at eSentire now has a mechanism in place to export Suricata alerts from esNetwork sensors and ingest them into the raw telemetry pipeline \cite{a2}. The alert stream tends to be very noisy with a high false positive rate.  The objective of this project is to develop a machine learning model that can classify incoming alerts as high likelihood vs. low likelihood for the purpose of noise reduction, and only escalate the most likely malicious activity to the attention of SOC analysts.
 
 \subsection{Project General Steps}
 
  To achieve the goal, this project was conducted by following general steps:
 \begin{enumerate}
     \item Getting access to the dataset and starting initially observing and understanding the raw data.
     \item Doing research on existing relevant approaches in articles and blogs.
     \item Processing data to extract potential features.
     \item Building AI models.
     \item Getting training, validating and testing results.
     \item Selecting a best-performed model as the primary result of this project.
 \end{enumerate}
Repeating 2-5 steps to adjust feature selection, model types, and model parameters to obtain a better result.

The rest of this article is organized as follows: Section \ref{rw} shows the previous research that has been done to reduce IDS/IPS false alert rate and some related neural network algorithms that were developed to classify incoming IDS/IPS signals. Section \ref{d} gives the ground truth of dataset pre-processing and tells the story of facing obstacles when labeling data. Section \ref{f} illustrates the strategy used for extracting the feature vector from original raw records. Section \ref{m} provides details on machine learning model selection and building. The main part is Section \ref{e}, which explains the experiment that has been done to verify the hypothesis step by step. And also gives the result, evaluation, and discussion. Section \ref{cf}
stresses the challenges of this project and points out the future exploring directions. Section \ref{c} draws the conclusion.

\section{Related Works}\label{rw}

Research on solving the false positive IDS/IPS alerts problem started about a  decade ago. In 2007,  Alshammari et al \cite{2} demonstrated a machine learning-based approach. It adopted neural networks and fuzzy logic and achieved an excellent accuracy of  90.92\%  by experimenting with  DARPA  1999  dataset.  But this scheme requires a considerable amount of training data. In 2010, Spathoulas and  Katsikas \cite{3} presented a  method to reduce the noise of the signature-based  IDS.  In their approach,  the similarity and the frequency of a  signature that triggers false alerts were considered as the feature to distinguish between false alerts and normal ones.  The filter they developed was constructed by three components namely the Neighboring Related Alerts (NRA) component, the High Alert Frequency (HAF) component and the Usual False Positives (UFP) component. Although the approach didn't involve machine learning, their work can reduce up to 75\% false alerts. In 2012, a semi-supervised learning model was shown by  Zhang and  Mei \cite{4}. Their biggest contribution was reducing the amount of training data required by supervised learning algorithms. However, at the same time, the false-negative rate was higher than in compared methods. In 2013, the Filter Parallelization method \cite{5} was presented. Researchers experimented with the KDD99 dataset, and they found out that SVM achieved the highest accuracy of 94.97\%, in comparison to KNN's accuracy (60.75\%). With the increasing diversity of the  Internet,  IDS/IPS  systems were implemented into industrial  IoT and  CPS (Cyber-Physical  Systems)  networks.  One of the recent research projects proposed a  machine learning-based method to reduce false alerts for IoT and CPS \cite{6}. This approach can self-modify the preventive rules according to the labelled samples received. The rules would be updated once a false-positive alert occurs to avoid false alerts. Besides, researchers have developed their z-classifiers with the CART classifier as the core. According to their experiment, this approach can achieve a zero false-positive rate. 

Despite the IDS false positive reducing approaches, some articles that present AI-based IDS/IPS systems can also provide some insights for this project. Shah and Trivedi \cite{7} have done a survey to show how neural networks can be leveraged for IDS/IPS systems. They compared different kinds of NNs that can be leveraged, such as Artificial Neural Network (ANN), and Back Propagation Neural Network (BPNN). Vollmer and Manic \cite{8} presented an IDS-NNM for critical infrastructures since they are used to control physical functions which are not immune to the threat of cyber attacks and may be potentially vulnerable. Karatas and Sahingoz \cite{9} did research on how model selection and training methods can impact the classification result. This approach can guide this project on model selection. Azizjon et al \cite{10} came up with several CNN structures that can classify TP/FP for an IDS system. According to their experiment, this approach has reached 90\% on average accuracy.

In \cite{a1}, Yazdinejad et al. proposed Kangaroo-based IDS for attack detection in software-defined networks (SDN). Indeed, this is an SDN-based architecture for attack detection and malicious behaviors in the data plane to solve security issues in the SDN. Similarly, I can mention some blockchain-based recent works \cite{a3,a4,a5,a6}, that applied blockchain and machine learning models to consider IDS and
decentralized environments. Also, these works focus on providing more security in authentication, threat detection, and hunting process with accurate machine learning models.
In another work \cite{c1}, proposed filtering out false positive grey matter atrophies in single subject voxel-based morphometry. Indeed, this is a machine learning technique widely used for automated data classification, namely Support Vector Machine (SVM), to refine the findings produced by SS-VBM. Also, in \cite{c2} Tiwari et al. considered Zeek IDS refinements that troubleshot and resolved some of the problems with the Zeek network monitoring tool. The logging capabilities of Zeek are enhanced by logging the City and Country names of IPs, filtering the noise of specific domains and ports, and separating local and remote connection Logs.

\section{Dataset}\label{d}

The data used in this work consists of the raw alert records generated by eSentire Atlas XDR Platform\footnote{https://www.esentire.com/what-we-do/extended-detection-and-response}. The Atlas continuously ingests alert signals from IDS cores, such as Suricata. It produces enriched raw alert records with analytical and expanded information.

\subsection{Data Acquisition}

\begin{figure}[htbp]
\centerline{\includegraphics[scale=0.7]{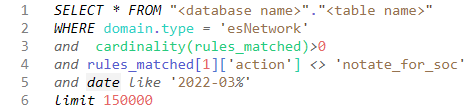}}
\caption{Sample Query used for scanning data.}
\label{getdata}
\end{figure}

The enriched alert raw records are stored in a database in Athena. To obtain the data, an SQL Query was created to scan and select data from massive raw records. In the SQL Query \cite{b1}, specific conditions were set to match the data I needed, and a time window was set to limit the amount of data. The SQL Query sample used to get data is shown in Figure \ref{getdata}. 

As you can see, the Query used to match data generated by $esNetwork$, at least matched 1 Atlas filter, is not targeting a specific client. And the date that the data was generated is March 2022.

Data that has been scanned and selected will be put into a datasheet automatically by Athena. Each row contains one single alert signal with fields including $raw\_record$, $enrichments$, and other metadata.

\subsection{Data Labeling}\label{A}

In my previous experience, there is always an open-source labelled dataset that can be leveraged to train and test a model directly. However, in terms of industrial workflows and pipelines, there is no such labelled dataset that I can directly use. Thus, I have to figure out a way to label the data with the help of specialists in my team as well as several SOC analysts. The data acquisition and labelling occupied most of the time during my project. 

\begin{figure}[htbp]
\centerline{\includegraphics[scale=0.35]{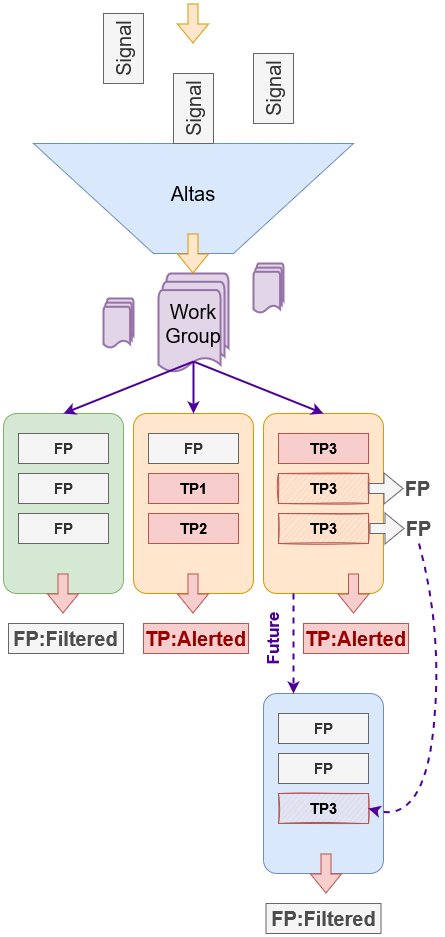}}
\caption{How Atlas handles signals.}
\label{atlas}
\end{figure}

According to one of the SOC analysts, the function of Atlas could be a little bit tricky. And the design of it leads to a lot of bad outcomes. Figure \ref{atlas} illustrates the way the Atlas deals with signals. As shown in the figure, the Atlas consume individual signals from IDS and randomly groups them into several batches called workgroups. Then the Atlas filters are applied to each workgroup to determine if they should be alerted or not. As long as a true positive signal was included in a workgroup, the entire group of signals will be marked as true positive (TP) \cite{a7}. If a workgroup contains duplicated TP signals, only one of them will be marked as TP \cite{a7}, the rest of them will be considered as false positive (FP) \cite{a7} and be filtered since they don't need to be alerted a second time. The problem is that Atlas makes a decision based on a workgroup instead of an individual signal. The conclusion drawn on a workgroup cannot be used to apply signals in it. As shown in the figure, a TP work group may contain several FP signals; an FP work group may have some previously alerted TP signals in it. This mechanism put the data labelling work in a dilemma.

\section{Feature extraction and selection}\label{f}

With the labelled raw data finally obtained, feature vectors can be extracted. Raw signal data consists of hundreds of fields that contain detailed information. In order to train the AI model, I extracted features from raw data into a vector so that they can clearly reveal the behaviours of events. After extraction, I selected the most useful features to feed the AI model.

\subsection{Feature extraction}

There are tons of information such as network transport details (like IP address), IDS system information (such as rules matched), preliminary analysis of the Atlas (such as event class type), and metadata(such as timestamps) \cite{a8}. Those details are stored with JSON strings. The next step is to convert the structured strings to a vector consisting of numerical values that machine learning models can understand.

Without knowing which field is useful, my first approach is to choose basic network packet information (i.e., IP addresses, port numbers, application layer details, etc.), alert signatures (i.e. rule ID number, rule descriptions, rule categories, etc.) based on my understanding of how IDS works and what could be critical elements for hunting a cyber attack.

Secondly, I obtained a small sample of data (approximately 1,000 signals), then used a Python script to parse the chosen fields and stored the result in a datasheet. Calculate each feature's variance and its Pearson correlation with the label to eliminate some low-variance and low-correlation features \cite{a9}.

\begin{figure}[htbp]
\centerline{\includegraphics[scale=0.5]{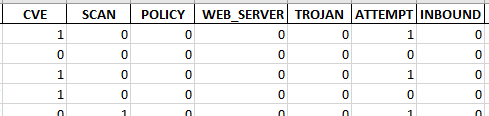}}
\caption{Features organized by One-Hot Encoding.}
\label{onehot}
\end{figure}

\begin{figure}[htbp]
\centerline{\includegraphics[scale=0.5]{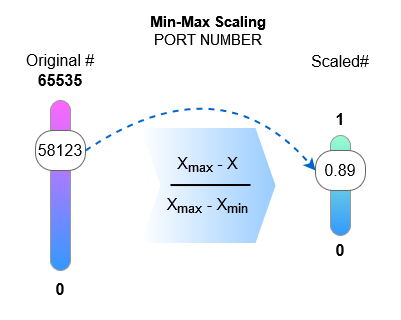}}
\caption{Apply Min-Max scaling to port number.}
\label{minmax}
\end{figure}

Finally, encoding the features using keyword searching, One-Hot Encoding, and Min-Max Scaling techniques. Keyword searching is performed to check the existence of specific keywords in string values. On-Hot Encoding is leveraged to organize the result of keyword searching by a vector. The sample data interpreted with these two methods are shown in Figure \ref{onehot}. Besides, machine learning models, mostly regression algorithms, do not perform well when numerical features are in a variety of ranges as models may consider large numbers as more important. Thus, I used Min-Max Scaling to convert each feature into the same scale. An example of applying Min-Max Scaling on port numbers is illustrated in Figure \ref{minmax}. As we all know, the minimum value of a valid port is 0, while the maximum is 65,535. This scaling method categorized the port number into 100 classes from 0-1. Applying scaling like this on features such as IP address, rule sid, and payload length can help maintain distinction without causing overfitting.

\subsection{Feature selection}

When the encoded features are ready, roughly the next step is to select the features. The feature selection took place at two stages--at the beginning of model training and the whole process later on.

Before training models, I tried to use the feature selection module by sklearn to select the top 30 features that are most likely to be helpful. The library $SelectKBest$ is what I adopted. It helps calculate each feature's $\chi \textsuperscript{2}$ score and select the top $k$ features.

After choosing models, the second stage of feature selection should be performed depending on each different AI model. During the training process, features were dynamically evaluated and re-selected. For instance, new features would be added to consideration. On this occasion, feature selection was repeatedly performed in each iteration. To verify the candidates during the model training, the SHAP \cite{b2,a10} value was calculated, which we will discuss in Section \ref{shapsec}.

\section{Machine Learning Models}\label{m}

In terms of machine learning models, I got several options to choose from. For instance, the models I am familiar with, such as Random Forest (RF), Support Vector Machine (SVM), and Multi-Layer Perceptron (MLP) \cite{b3}. Besides, what I can think about is Neural Networks \cite{b4}. 

To select the top-performing ones, I used the LazyClassifier library by LazyPredict to do my experiment.

\subsection{Exploring potential models}

\begin{figure}[htbp]
\centerline{\includegraphics[scale=0.7]{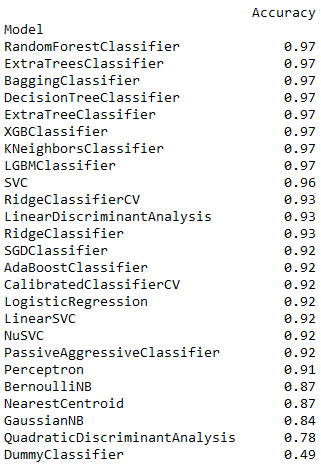}}
\caption{Result of LazyClassifier.}
\label{lazy}
\end{figure}

The LazyClassifier accepts input data and performs training with 30 standard models \cite{b5}. Then it can sort the models by the result. All I need to do is pick the top ones and continue exploring with them. In the first stage, I used the LazyClassifier with pre-processed data. Figure \ref{lazy} shows the result I got.

From the figure, I can tell that trees and forests have the highest potential, and so does SVM (i.e., SVC). So, the second stage is to compare my choices. By comparing SVC and RF, I found out RF is my top choice because it got 97\% of the training result while SVC only got 91\%. However, both were overfitting somehow due to the model without fine-tuning and the dataset without further optimization. While the LazyClassifier does not support neural networks, I decided to include RF, MLP, and 1D-CNN in my selection and do a further experiment.

\section{Experiment}\label{e}

This experiment consists of several cycles, which are very similar to PDCA\footnote{https://en.wikipedia.org/wiki/PDCA} cycles. Moreover, this is the task I spent significant time on. 

Each cycle can be described as checking the training and validating result, adjusting the features selected (or adding new features that may help), changing the model's parameters, checking if the dataset needs to be modified, and obtaining the result.
%
%\begin{figure}[htbp]
%\centerline{\includegraphics[scale=0.35]{pdca.png}}
%\caption{One experiment cycle.}
%\label{pdca}
%\end{figure}

\subsection{Data labeling}\label{labelingsec}

\begin{figure}[htbp]
\centerline{\includegraphics[scale=0.4]{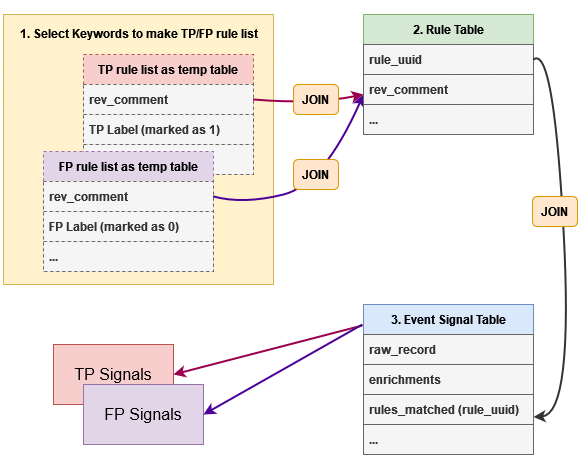}}
\caption{Join 3 tables to label data.}
\label{labeling}
\end{figure}

With the problem mentioned in section \ref{A}, A SOC analyst offered a brilliant idea that I can label signals by referring to the SOC's comments on the filter. The procedure was described in Figure \ref{labeling}. For short, some of the Atlas filters contain comments given by analysts. Those comments are stored in the rule table and the corresponding rule $uuid$. I can label the signals manually by joining the signal table, rule table, and rule lists which mark the comments as TP/FP. For instance, if a filter rule was commented as "filtering benign activity, " every signal filtered by it can be considered FP. Signals went through the filter commented as "External scan has been alerted" can be labeled as TP.

This approach was a piece of music to my ear. However, it is still defective in some situations. Firstly, tons of comments say TP/FP, but they don't provide information about the reasons. Secondly, the comment can sometimes be very vague. Analysts wrote very long sentences to describe the situation without concluding. Thirdly, some of the comments are conditional. For instance, the comment tells that the signal should be confirmed by SOC only if the source IP can be trusted. However, on my side, there is no way to find out if the source IP is trusted or not.

Failure to accurately label data can cause a severe consequence of low accuracy because the AI model can be easily confused \cite{a11}. To be safe than sorry, I had to ignore those ambiguously commented filters. 

With that, here comes another hurdle. After I abandoned those rules with ambiguous comments, I can only get a very limited number of signals. For example, when I obtain approximately 10,000 signals, only about 10\% of them can be successfully labeled. The situation is even worse regarding TP labeling, as true positive alters occupy only a very small portion of total IDS alerts. To obtain enough data, I had to go through TBs of data in Athena. Nevertheless, due to the special mechanism of the Atlas, I do not have any other choice. My team and I believe that is the most feasible approach so far. So, the labeled datasets used in this work are not very promising.

Based on the idea, there are two ways to perform labelling:
\begin{itemize}
    \item Download raw data and match it with TP/FP rule $uuid$ lists using Python script.
    \item Label data while scanning using SQL Query.
\end{itemize}

The former method was adopted in the first two months when I were still getting used to the Athena. Firstly, TP/FP filter's rule $uuid$ lists were obtained based on the keyword search. For example, any comments contains $'expected'$, $'benign'$, $'whitelisted'$... should be put into FP list; comments contains $'alerted'$, $'sent'$... will go into TP list. Thus, TP/FP lists contain only two column: $rule\_uuid$ and $rev\_comment$.
I downloaded raw data from Athena and used a Python script to match the raw record data by the $uuid$ in TP/FP lists. 

%\begin{figure}[htbp]
%\centerline{\includegraphics[scale=0.5]{tplist.PNG}}
%\caption{Query for generating TP rule list.}
%\label{tplist}
%\end{figure}

%\begin{figure}[htbp]
%\centerline{\includegraphics[scale=0.5]{fplist.PNG}}
%\caption{Query for generating FP rule list.}
%\label{fplist}
%\end{figure}

This method's advantage is that the lists' comments are entirely audible. However, the pain is that I had to download TBs of raw data to my local machine and upload it to the Jupyter Notebook online. It is a very time-consuming job and basically with very low efficiency.

I devised another way to label data to save time for other jobs. This way is an improved solution, which can obtain the lists and match data by Athena and the process that scans data from the database.

The latter method is to add a $WITH$ clause on top of the SQL Query to get the TP/FP lists in the same way mentioned above, and use it as a temporary table to join on the events table, as the way shown in Figure \ref{labeling} so that I can label the data at the same time. It saved a huge amount of time because there is no need to download all the data while eventually, just 10\% of them can be successfully labeled. 

However, besides that, data that is used to train and test should be further selected. Data that has $action = notate\_for\_soc$ doesn't represent general security events. Instead, those rules only applied to clients wide. Thus, in the $Line 4$ of Figure \ref{getdata}, $action <> notate\_for\_soc$ has been applied as a condition to filter out that dat

\subsection{Training data preparation}\label{training}

\begin{figure}[htbp]
\centerline{\includegraphics[scale=0.58]{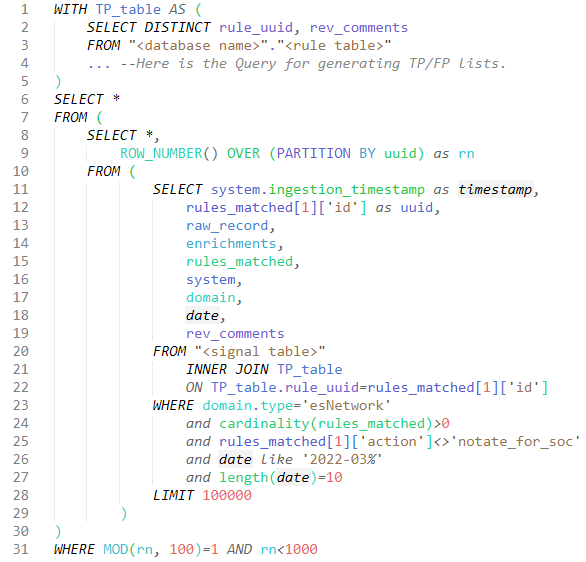}}
\caption{Optimized Query for labeling and scanning.}
\label{label22}
\end{figure}

Training data should be able to represent as many kinds of events as possible so that the model trained with it can cope with more complicated situations. However, data obtained using the SQL query in section \ref{A} is highly duplicated. For instance, if an internal network port scanning happened on the day I chose, the 10,000 lines of signals I get could be generated from only 1 sensor for the same reason. The only thing different is the port number. If I use those 10,000 lines of data for training, that's not making any sense. Thus, the optimized SQL Query was provided in Figure \ref{label22}. In addition to scan data, I partitioned the data by the rule id ($uuid$) that they matched. Then as you can see $Line 31$ in the figure, I only choose the first item in every 100 items and no more than 10 items that matched by the same rule. This turns out to be very helpful since it can get rid of highly duplicated signals while still allowing a very slight duplication. This will help AI models learn further from each feature in the vector.

After iterating 8 times, the latest version of the training dataset includes 982 TP samples and 1,126 FP samples covering from January 2022 to April 2022. 

\subsection{Testing data preparation}

The same procedure can be applied to obtaining the testing dataset. The only change that should be made is to entirely avoid duplication so that I can calculate accurate precision and recall scores. My testing dataset covers 210 TPs and 251 FPs from May 2022 to July 2022, completely isolated from the training dataset.

In addition to that, a small set of data that a prestigious Tire 3 SOC analyst reviewed was adopted to test the performance. Each signal was commented on whether it should be escalated to an analyst with the reason behind that given.

\subsection{Feature Selection}

With adjusting features selected during each cycle of the experiment, the final version of the feature vector includes the items explained in Table \ref{features}.

\begin{table*}[htbp]
\caption{Features in use}
\begin{center}
\begin{tabular}{|p{1.5em}|p{9em}|p{5.5cm}|p{5cm}|}
\hline
\textbf{ }&\multicolumn{3}{|c|}{\textbf{Properties}} \\
\cline{2-4} 
\textbf{\#} & \textbf{\textit{Name in script}}& \textbf{\textit{Description}}& \textbf{\textit{Type\&Range}} \\
\hline
1&priv\_src\_ip&Private Src IP?&Boolean (1=YES, 0=NO)\\
2&priv\_dst\_ip&Private Dest IP?&Boolean (1=YES, 0=NO)\\
3&sip&Src IP (Re-scaled)&Float64 (0.00--1.00)\\
4&dip&Dest IP (Re-scaled)&Float64 (0.00--1.00)\\
5&diff&Src/Dest IP Difference&Float64 (0.00--1.00)\\
6&http\_status&HTTP status code&Float64 (0.000, 0.2xx, 0.3xx, 0.4xx, 0.5xx)\\
7&pkt\_to\_svr&\# of packets to server&Float64 (0.00--1.00, -1.00)\\
8&pkt\_to\_clt&\# of packets to client&Float64 (0.00--1.00, -1.00)\\
9&byt\_to\_svr&\# of bytes to server&Float64 (0.00--1.00, -1.00)\\
10&byt\_to\_clt&\# of bytes to client&Float64 (0.00--1.00, -1.00)\\
11&rulesid&Rule SID&Float64 (0.00--1.00)\\
12&CVE&Rule description contains 'CVE'?&Boolean (1=YES, 0=NO)\\
13&attack&Class type contains 'attack'?&Boolean (1=Yes, 0=NO)\\
14&EXPLOIT&Rule description contains 'EXPLOIT'?&Boolean (1=Yes, 0=NO)\\
15&POSSIBLE&Rule description contains 'POSSIBLE'?&Boolean (1=Yes, 0=NO)\\
16&activity&Class type contains 'activity'?&Boolean (1=Yes, 0=NO)\\
17&attempt&Class type contains 'attempt'?&Boolean (1=Yes, 0=NO)\\
18&sport&Src Port (Re-scaled)&Float64 (0.00--1.00)\\
19&dport&Dest Port (Re-scaled)&Float64 (0.00--1.00)\\
20&PAYLOAD\_Bytes&Payload length in bytes&Float64 (0.00--1.00)\\
*21&SCAN&Rule description contains 'SCAN'?&Boolean (1=Yes, 0=NO)\\
*22&POLICY&Rule description contains 'POLICY'?&Boolean (1=Yes, 0=NO)\\
*23&WEB\_SERVER&Rule description contains 'WEB\_SERVER'?&Boolean (1=Yes, 0=NO)\\
*24&TROJAN&Rule description contains 'TROJAN'?&Boolean (1=Yes, 0=NO)\\
*25&ATTEMPT&Rule description contains 'ATTEMPT'?&Boolean (1=Yes, 0=NO)\\
*26&INBOUND&Rule description contains 'INBOUND'?&Boolean (1=Yes, 0=NO)\\
*27&UNUSUAL&Rule description contains 'UNUSUAL'?&Boolean (1=Yes, 0=NO)\\
*28&.(dot)&Rule description contains '.'?&Boolean (1=Yes, 0=NO)\\
*29&policy&Class type contains 'policy'?&Boolean (1=Yes, 0=NO)\\

\hline
\multicolumn{4}{l}{*Features are only used for 1-DCNN model.}
\end{tabular}
\label{features}
\end{center}
\end{table*}

When it comes to RF and MLP, there is no need to use them all. Those features with an asterisk (*) at the beginning of the feature number will be excluded for those two models and only appears in 1D-CNN.

\subsection{Model Training and testing}\label{shapsec}
\begin{figure}[htbp]
\centerline{\includegraphics[scale=0.6]{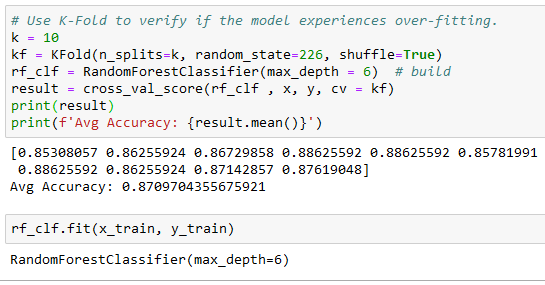}}
\caption{Code for building the model.}
\label{rf}
\end{figure}

The Random Forest Classifier was built by $sklearn. Ensemble. RandomForestClassifier$ accepts customized parameters such as the number of estimators and the max depth of the forest. Initially, I didn't customize any parameters, so the default number of estimators is 100, and there is no max depth limit set by default. This has caused an overfitting problem. After fixing the problem, I found out $RandomForestClassifier(n\_estimators = 100, max\_depth = 6)$ provided the best performance. The Python code used for building and fitting is shown in Figure \ref{rf}. The training TP recall is 93\%, FP recall is 77\%, with an accuracy of 85\%.

\begin{figure}[htbp]
\centerline{\includegraphics[scale=0.4]{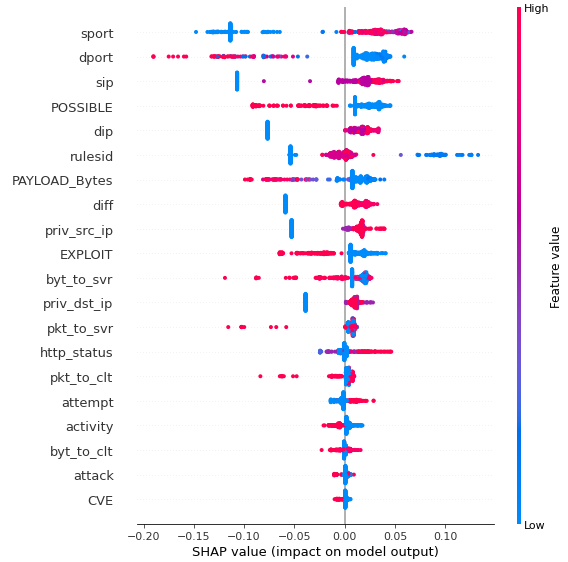}}
\caption{SHAP value for each feature.}
\label{shap1}
\end{figure}

\begin{figure}[htbp]
\centerline{\includegraphics[scale=0.4]{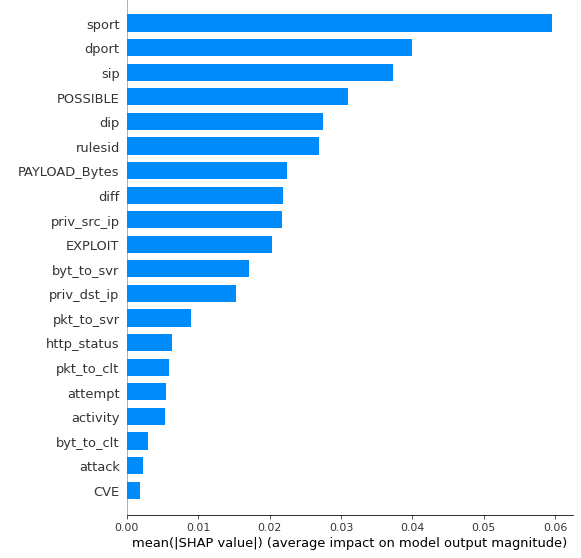}}
\caption{Contributions made by each feature.}
\label{shap2}
\end{figure}

The $max\_depth$ tells the model how deep it allows to learn. The deeper it goes, the higher the overfitting likelihood will be. The SHAP value was adopted each time I got the training result from the RF model. Figure \ref{shap1} demonstrates the latest SHAP value for each feature. The greater the degree of dispersion, the more contribution was made by the feature. Figure \ref{shap2} shows the contribution of each feature sorted from high to low. To adjust the features after each training, I removed the features shown at the tail and substituted them with new features. 

The other two alternative models have not been fully developed due to the time limit of this project. However, both have huge potential to achieve the goal better. The following is how they were built:

\begin{itemize}
    \item MLP: $sklearn.neural\_network.MLPClassifier$ with the default setting for parameters.
    \item 1D-CNN: This was inspired by research for CNN-based intrusion detection system by Meliboev et al \cite{1dcnn}. What this approach does is very similar to my project with the objective--Classifying incoming signals into two categories. So, I've tried to create a 3-layer CNN model described in the Figure \ref{cnn} following their findings.
    
\begin{figure}[htbp]
\centerline{\includegraphics[scale=0.58]{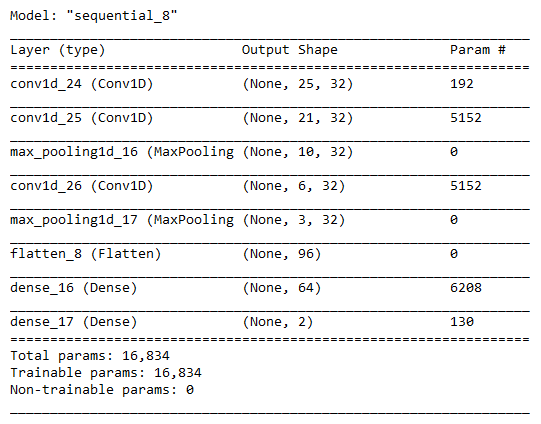}}
\caption{3-Layer 1-DCNN model structure.}
\label{cnn}
\end{figure}

\end{itemize}

\subsection{Problem: Overfitting}

In my first attempt to build the random forest classifier, I got a training validation accuracy of 98\% while the testing accuracy was 80\%. Every experienced team member said it was probably overfitting. So, I looked for the cause of overfitting and the common solutions. At that time, the model was premature, and the features were not well-encoded.

I used $k$-Fold validation and re-scaled some of the features to solve the problem. $k$-Fold is a scheme that can divide the training set into $k$ parts, choose one part as the validation set and other parts to be the training set, test the model for $k$ times, and display the accuracy of each time. If some of the parts got an extraordinary or pretty poor accuracy, it indicates the model might be overfitting.

The 10-Fold test had a result with high-variance accuracy scores, which indicated the model was overfitting. Then I checked my features. Some of the float-type features are not in fixed length. For instance, the rule $sid$ could be drastically different. At that time, I didn't classify ids into several fixed categories. I kept them the way they were, like 0.2102234, 0.08976, instead of rounding the float number to reserve 1 decimal place (i.e., 0.0-1.0). This mistake confused the model so that when it comes to a new $sid$, the model does not know what should be done.

To reduce the overfitting, I reserved 3 decimal places for those drastically various numbers. That means categorizing those numbers into 1,001 classes (i.e., 0.000-1.000). So, any number from new signals can correspond to a category.

\subsection{Result}

After two months of fine-tuning and experiments, the final result of the RF model, the preliminary result for MLP and 1D-CNN are displayed in Table \ref{result}.

\begin{table}[htbp]
\caption{Experiment Result}
\begin{center}
\begin{tabular}{|c||c|c|c|c|c|}
\hline
\textbf{Model}&\multicolumn{2}{|c|}{\textbf{TP}}&\multicolumn{2}{|c|}{\textbf{FP}}&\textbf{Accuracy} \\
\cline{2-5} 

&  \textbf{Precision} & \textbf{Recall} & \textbf{Precision} & \textbf{Recall} & \\
\hline
\textbf{RF} & \textbf{83\%} & \textbf{98\%} & \textbf{98\%} & \textbf{83\%} & \textbf{90\%}\\
MLP& 86\% & 95\% & 96\% & 87\% & 91\%\\
1D-CNN& 85\% & 93\% & 94\% & 86\% & 89\%\\
\hline

\end{tabular}
\label{result}
\end{center}
\end{table}

Generally, the accuracy scores of the RF and the MLP are better than the 1D-CNN. The reason could be the underdevelopment due to the limited time. In terms of RF and MLP, both of them achieved acceptable accuracy. But RF showed a higher TP recall (98\%), which means it can successfully catch more true positive alerts.

\subsection{Evaluation}

Although numerous approaches have dealt with IDS false positive alerts that can achieve nearly perfect accuracy when it comes to industrial application, situations could be drastically different and more complex \cite{a12}. Therefore, high accuracy should not be the only criterion for evaluation as it has always been in academic research. For instance, a high TP recall is more critical for filtering alerts from IDS since it is unacceptable if my model accidentally filters a significant number of true positive alerts. Catching as many TP signals as possible should prioritize more than filtering as many FP alerts. Because of this, I chose the Random Forest Classifier, which was able to catch 98\% TP alerts, as my priority approach for this project.

To further simulate the real situation, we obtained a list of security event signals that escalated to SOC on July 28, 2022, grouped by $investigation link$. This list contains 65 TP and 327 FP events, which represents the total alerts that the SOC team worked on. Using it as a test set, the result shows the RF model can filter out 285 out of 327 false positive alters (which is 87\%) without mistakenly muting one single real alert. 
 
Due to the difficulty of getting the data labeled reliably, the accuracy of these classifiers is moderate. However, it does not mean the model is defective. There is so much to explore in the future. The MLP and 1D-CNN models have a high potential to perform better. Besides, there are so many potential features that can be added to consideration. 

\section{Project Challenges and Future Work}\label{cf}
There are several obstacles and challenges in this project that I have discussed above. Thus, a summary is given to stress and underlines the hardship together with future works to improve.
\begin{enumerate}
    \item Lacking of reliable labeling scheme. As I discussed in Section \ref{A}, I had to join three tables to search and label TP/FP signals based on the rev\_comments they have. The feasibility of this method should be based on the reliability of the comments. However, those comments were drawn by different SOC analysts. They all had different understandings and insights about the signals. Hence, I cannot get precisely labeled data from those comments.
    
    To further lift the performance of my models in the future, one of the critical problems to solve is to figure out other more promising ways to label the data. 
    
    \item Data deficiency. The database itself is full of various data generated day by day. Nevertheless, what actually caused the deficiency is the following problems. As I mentioned in Section \ref{labelingsec} and \ref{training}, signals stored in the database are highly duplicated. Besides, data only has the field $action <> notate\_for\_soc$ that can be used for training and testing. In this situation, I had to use SQL Query to scan several TBs of signals and exclude about 90\% of what I scanned to get only a thousand lines of signals.

\end{enumerate}

Due to this project's time limit, much meaningful work can be done in the future. For instance, optimize the datasets used for training and testing by upgrading the labeling method. Besides, models adopted in this project can be further developed and fine-tuned to fit the data more appropriately. Last but not least, the model with an ideal result can be put into industrial pipelines as a prototype to verify its performance with dynamic workflows. Once the prototype is proved efficient, it can be deployed to serve the industry.

\section{Conclusion}\label{c}

The project objective, which is to filter out esNetwork false-positive alerts, was successfully achieved, even though a few details can still be further fine-tuned.

As explained by the experiment results, three machine learning models involved in this research (RF, MLP, 1D-CNN) showed their significant contribution to filtering false positive alerts that were produced by the esNetwork. Among all three models, although the overall accuracy is just 90\%, RF performed best as it can achieve the highest true positive recall, which is 98\%. The experiment successfully filtered out 208 FP alerts out of 251, which is estimated to help the SOC team save 13.8 hours of working time. According to the simulation test with real industry pipeline data, the RF model filtered out 285 FP alerts out of 327 on July 28, 2022, saving approximately 20 hours of work for the SOC team. The estimated cost deduction for the corporation is 208,000 Canadian dollars per year.

\section*{Acknowledgment}

I would like to thank people who have been supportive throughout my project. 

I am very grateful to my professor Dr. Ali Dehghantanha and my supervisor Mr. Wafic Al-Wazzan at eSentire Inc. for their non-stop support. A special thank you of mine goes to my colleagues in the ATA team who helped me out in completing this internship project, where they all exchanged their brilliant ideas and made it possible for me to complete the project with all accurate information.

Finally, thanks to the University of Guelph and eSentire Inc. for offering me such a precious opportunity to leverage what I've learned to an enterprise stage and further develop my professional skills.

\bibliographystyle{ieeetr}
\bibliography{ref2}

\begin{thebibliography}{10}

\bibitem{a1}
A.~Yazdinejadna, R.~M. Parizi, A.~Dehghantanha, and M.~S. Khan, ``A
  kangaroo-based intrusion detection system on software-defined networks,''
  {\em Computer Networks}, vol.~184, p.~107688, 2021.

\bibitem{1}
T.~Pietraszek and A.~Tanner, ``Data mining and machine learning—towards
  reducing false positives in intrusion detection,'' {\em Information security
  technical report}, vol.~10, no.~3, pp.~169--183, 2005.

\bibitem{a2}
A.~Yazdinejad, A.~Dehghantanha, R.~M. Parizi, M.~Hammoudeh, H.~Karimipour, and
  G.~Srivastava, ``Block hunter: Federated learning for cyber threat hunting in
  blockchain-based iiot networks,'' {\em IEEE Transactions on Industrial
  Informatics}, pp.~1--1, 2022.

\bibitem{2}
R.~Alshammari, S.~Sonamthiang, M.~Teimouri, and D.~Riordan, ``Using neuro-fuzzy
  approach to reduce false positive alerts,'' in {\em Fifth Annual Conference
  on Communication Networks and Services Research (CNSR'07)}, pp.~345--349,
  IEEE, 2007.

\bibitem{3}
G.~P. Spathoulas and S.~K. Katsikas, ``Reducing false positives in intrusion
  detection systems,'' {\em computers \& security}, vol.~29, no.~1, pp.~35--44,
  2010.

\bibitem{4}
M.~Zhang and H.~Mei, ``A new method for filtering ids false positives with
  semi-supervised classification,'' in {\em International Conference on
  Intelligent Computing}, pp.~513--519, Springer, 2012.

\bibitem{5}
M.~Amanifar and M.~Saniee~Abadeh, ``Accuracy improvement of ids via filter
  parallelization.,'' {\em International Journal of Advanced Research in
  Computer Science}, vol.~4, no.~10, 2013.

\bibitem{6}
M.~S. Haghighi, F.~Farivar, and A.~Jolfaei, ``A machine learning-based approach
  to build zero false-positive ipss for industrial iot and cps with a case
  study on power grids security,'' {\em IEEE Transactions on Industry
  Applications}, 2020.

\bibitem{7}
B.~Shah and B.~H. Trivedi, ``Artificial neural network based intrusion
  detection system: A survey,'' {\em International Journal of Computer
  Applications}, vol.~39, no.~6, pp.~13--18, 2012.

\bibitem{8}
O.~Linda, T.~Vollmer, and M.~Manic, ``Neural network based intrusion detection
  system for critical infrastructures,'' in {\em 2009 international joint
  conference on neural networks}, pp.~1827--1834, IEEE, 2009.

\bibitem{9}
G.~Karatas and O.~K. Sahingoz, ``Neural network based intrusion detection
  systems with different training functions,'' in {\em 2018 6th International
  Symposium on Digital Forensic and Security (ISDFS)}, pp.~1--6, IEEE, 2018.

\bibitem{10}
G.~Karatas and O.~K. Sahingoz, ``Neural network based intrusion detection
  systems with different training functions,'' in {\em 2018 6th International
  Symposium on Digital Forensic and Security (ISDFS)}, pp.~1--6, IEEE, 2018.

\bibitem{a3}
A.~Yazdinejad, R.~M. Parizi, A.~Dehghantanha, Q.~Zhang, and K.-K.~R. Choo, ``An
  energy-efficient sdn controller architecture for iot networks with
  blockchain-based security,'' {\em IEEE Transactions on Services Computing},
  vol.~13, no.~4, pp.~625--638, 2020.

\bibitem{a4}
A.~Yazdinejad, G.~Srivastava, R.~M. Parizi, A.~Dehghantanha, K.-K.~R. Choo, and
  M.~Aledhari, ``Decentralized authentication of distributed patients in
  hospital networks using blockchain,'' {\em IEEE journal of biomedical and
  health informatics}, vol.~24, no.~8, pp.~2146--2156, 2020.

\bibitem{a5}
A.~Yazdinejad, R.~M. Parizi, A.~Dehghantanha, and K.-K.~R. Choo,
  ``Blockchain-enabled authentication handover with efficient privacy
  protection in sdn-based 5g networks,'' {\em IEEE Transactions on Network
  Science and Engineering}, vol.~8, no.~2, pp.~1120--1132, 2019.

\bibitem{a6}
A.~Yazdinejad, R.~M. Parizi, A.~Dehghantanha, H.~Karimipour, G.~Srivastava, and
  M.~Aledhari, ``Enabling drones in the internet of things with decentralized
  blockchain-based security,'' {\em IEEE Internet of Things Journal}, vol.~8,
  no.~8, pp.~6406--6415, 2020.

\bibitem{c1}
H.~C. K{\"u}lsgaard, J.~I. Orlando, M.~Bendersky, J.~P. Princich, L.~S.
  Manzanera, A.~Vargas, S.~Kochen, and I.~Larrabide, ``Machine learning for
  filtering out false positive grey matter atrophies in single subject voxel
  based morphometry: a simulation based study,'' {\em Journal of the
  Neurological Sciences}, vol.~420, p.~117220, 2021.

\bibitem{c2}
A.~Tiwari, S.~Saraswat, U.~Dixit, and S.~Pandey, ``Refinements in zeek
  intrusion detection system,'' in {\em 2022 8th International Conference on
  Advanced Computing and Communication Systems (ICACCS)}, vol.~1, pp.~974--979,
  2022.

\bibitem{b1}
R.~K. Dhanaraj, V.~Ramakrishnan, M.~Poongodi, L.~Krishnasamy, M.~Hamdi,
  K.~Kotecha, and V.~Vijayakumar, ``Random forest bagging and x-means clustered
  antipattern detection from sql query log for accessing secure mobile data,''
  {\em Wireless Communications and Mobile Computing}, vol.~2021, 2021.

\bibitem{a7}
A.~Yazdinejad, H.~HaddadPajouh, A.~Dehghantanha, R.~M. Parizi, G.~Srivastava,
  and M.-Y. Chen, ``Cryptocurrency malware hunting: A deep recurrent neural
  network approach,'' {\em Applied Soft Computing}, vol.~96, p.~106630, 2020.

\bibitem{a8}
A.~Yazdinejad, A.~Bohlooli, and K.~Jamshidi, ``Efficient design and hardware
  implementation of the openflow v1. 3 switch on the virtex-6 fpga ml605,''
  {\em The Journal of Supercomputing}, vol.~74, no.~3, pp.~1299--1320, 2018.

\bibitem{a9}
S.~Nakhodchi, B.~Zolfaghari, A.~Yazdinejad, and A.~Dehghantanha, ``Steeleye: An
  application-layer attack detection and attribution model in industrial
  control systems using semi-deep learning,'' in {\em 2021 18th International
  Conference on Privacy, Security and Trust (PST)}, pp.~1--8, IEEE, 2021.

\bibitem{b2}
K.~E. Mokhtari, B.~P. Higdon, and A.~Ba{\c{s}}ar, ``Interpreting financial time
  series with shap values,'' in {\em Proceedings of the 29th Annual
  International Conference on Computer Science and Software Engineering},
  pp.~166--172, 2019.

\bibitem{a10}
Y.~Hailemariam, A.~Yazdinejad, R.~M. Parizi, G.~Srivastava, and
  A.~Dehghantanha, ``An empirical evaluation of ai deep explainable tools,'' in
  {\em 2020 IEEE Globecom Workshops (GC Wkshps}, pp.~1--6, IEEE, 2020.

\bibitem{b3}
H.~Taud and J.~Mas, ``Multilayer perceptron (mlp),'' in {\em Geomatic
  approaches for modeling land change scenarios}, pp.~451--455, Springer, 2018.

\bibitem{b4}
W.~Samek, G.~Montavon, S.~Lapuschkin, C.~J. Anders, and K.-R. M{\"u}ller,
  ``Explaining deep neural networks and beyond: A review of methods and
  applications,'' {\em Proceedings of the IEEE}, vol.~109, no.~3, pp.~247--278,
  2021.

\bibitem{b5}
A.~Ghazal and S.~O. Kuznetsov, ``Lazy classification of underground forums
  messages using pattern structures,'' {\em FCA4AI 2022}, p.~63, 2022.

\bibitem{a11}
A.~Yazdinejad, R.~M. Parizi, A.~Dehghantanha, and H.~Karimipour, ``Federated
  learning for drone authentication,'' {\em Ad Hoc Networks}, vol.~120,
  p.~102574, 2021.

\bibitem{1dcnn}
M.~Azizjon, A.~Jumabek, and W.~Kim, ``1d cnn based network intrusion detection
  with normalization on imbalanced data,'' in {\em 2020 International
  Conference on Artificial Intelligence in Information and Communication
  (ICAIIC)}, pp.~218--224, IEEE, 2020.

\bibitem{a12}
A.~Yazdinejad, E.~Rabieinejad, A.~Dehghantanha, R.~M. Parizi, and
  G.~Srivastava, ``A machine learning-based sdn controller framework for drone
  management,'' in {\em 2021 IEEE Globecom Workshops (GC Wkshps)}, pp.~1--6,
  IEEE, 2021.

\end{thebibliography}

\end{document}